\documentclass{sig-alternate}

\usepackage{amsmath}
\usepackage{amssymb}
\usepackage{url}

\newcommand{\bC}{ {\mathbb C}}

\newcommand{\bQ} { {\mathbb Q}}

\newcommand{\cA}{ {\cal A}}

\newcommand{\cM}{ {\cal M}}

\newcommand{\LO} { {\mathcal L} }

\newcommand{\ld}{{\ell\delta}}
\newcommand{\ls}{{\ell\sigma}}
\newcommand{\lphi}{{\ell\phi}}
\newcommand{\lpsi}{{\ell\psi}}
\newcommand{\ve} { {\bf e}}

\newcommand{\pa} { \partial }
\newcommand{\s} { \sigma }

\newcommand{\vb} { {\bf b}}

\newcommand{\vu} { {\bf u}}
\newcommand{\vv}{ {\bf v}}
\newcommand{\vw} { {\bf w}}

\newtheorem{thm}{Theorem}
\newtheorem{cor}[thm]{Corollary}
\newtheorem{lemma}[thm]{Lemma}
\newtheorem{prop}[thm]{Proposition}
\newtheorem{defn}{Definition}
\newtheorem{ex}[defn]{Example}

\def\more-auths{%
\end{tabular}
\begin{tabular}{c}}
\begin{document}
%
\conferenceinfo{ISSAC}{'06, July 9-12, 2006, Genova, Italy }

\title{A Recursive Method for
Determining the \\ One-Dimensional Submodules of  Laurent-Ore
Modules
\titlenote{{ This research was supported in part by the National
Science Foundation of the USA under Grants CCR-0096842 (Singer) and
OISE-0456285 (Li, Singer, Zheng), and by a 973
project of China 2004CB31830 (Li, Wu, Zheng).}} }

\numberofauthors{2}

\author{
\alignauthor Ziming Li \\
        \affaddr{Key Lab of Mathematics-Mechanization} \\
        \affaddr{Academy of Mathematics and System Sciences} \\
        \affaddr{Zhong Guan Cun, Beijing 100080, China}\\
        \email{zmli@mmrc.iss.ac.cn}
\alignauthor Michael F. Singer \\
        \affaddr{Department of Mathematics, Box 8205}\\
        \affaddr{North Carolina State University}\\
        \affaddr{ Raleigh, NC, 27695-8205, USA}\\
        \email{singer@math.ncsu.edu}
\more-auths
\alignauthor Min Wu \\
        \affaddr{Software Engineering Institute}\\
        \affaddr{ East China Normal University}\\
        \affaddr{North Zhongshan Rd, Shanghai, 200062, China}\\
        \email{mwu@sei.ecnu.edu.cn}
\alignauthor Dabin Zheng \\
        \affaddr{Key Lab of Mathematics-Mechanization, Academy of Mathematics and
System Sciences} \\
        \affaddr{Zhong Guan Cun, Beijing 100080,
China} \\
        \email{zhengdabin@mmrc.iss.ac.cn}
}
\date{}
\maketitle

\begin{abstract}
We present a method for determining the one-dimensional submodules
of  a Laurent-Ore module. The method is based on a correspondence
between hyperexponential solutions of associated systems and
one-dimensional submodules. The hyperexponential solutions are
computed recursively by solving a sequence of first-order ordinary
matrix equations. As the recursion proceeds, the matrix equations
will have constant coefficients with respect to the operators that
have been considered.
\end{abstract}



\section{Introduction}

A Laurent-Ore algebra~$\LO$ over a field is a mathematical
abstraction of common properties of linear partial differential and
difference operators. Finite-dimensional $\LO$-modules interpret
finite-dimensional systems of linear partial differential and
difference equations concisely and precisely. For example, a factor
of a finite-dimensional system corresponds to a submodule of its
module of formal solutions~\cite{BronsteinLiWu05, Wu_thesis}. A
method for factoring finite-dimensional systems of linear PDE's is
given in~\cite{LiSchwarzTsarev03}, and, recently, a method for
factoring finite-dimensional~$\LO$-modules is presented
in~\cite{Wu_thesis}. Both are generalizations of the associated
equations method dated back to Beke~\cite{Beke1894}. A basic step in
these methods is to compute one-dimensional submodules of some
exterior powers of the given module. One approach for computing
one-dimensional submodules is to identify all possible partial
\lq\lq logarithmic derivatives\rq\rq of hyperexponential solutions
with respect to each differential or difference operator, and then,
glue the partial results together by common associates, as described
in~\cite{LabahnLi04, LiSchwarz01, LiSchwarzTsarev03}. In this
approach, one would have to compute hyperexponential solutions of
several ordinary differential and difference equations over the
ground field~$F$, and in addition, it is only applicable when~$F$ is
a field of rational functions and each operator acts on only one
variable non-trivially.

In this paper, we describe a method that is recursive on the set of
differential and difference operators acting on~$F$. It computes
hyperexponential solutions of an (ordinary) matrix equation and then
proceeds by back-substitution. In doing so, one avoids computing all
possible partial \lq\lq logarithmic derivatives\rq\rq of
hyperexponential solutions, which may be costly. Each time an
operator is carried on, we can reduce our problem to solving a
first-order matrix equation whose coefficients are constants in~$F$
with respect to the operator. {So the systems to be solved become
simpler as the computation goes on.
In particular, we are able to remove the restrictions imposed
in~\cite{LabahnLi04, LiSchwarz01, LiSchwarzTsarev03, Wu_thesis} on
operators and now require only that they commute.}

The rest of this paper is organized as follows. In
Section~\ref{SECT:pre}, we present some preliminaries and define the
notion of hyperexponential vectors. In Section~\ref{SECT:mlo}, we
describe a correspondence between the one-dimensional submodules of
an~$\LO$-module and the hyperexponential solutions of associated
systems. In Section~\ref{SECT:hsol}, we demonstrate how to identify
unspecified
 constants appearing in hyperexponential vectors to make these
vectors extensible. In Section~\ref{SECT:alg} we describe an
algorithm for determining the one-dimensional submodules of
an~$\LO$-module and give some examples.
\section{Preliminaries} \label{SECT:pre}
Throughout the paper,~$F$ is a commutative field of characteristic
zero.  Let~$R$ be an $F$-algebra, not necessarily commutative. We
present some basic facts about finite-dimensional $R$-modules, and
the notion of hyperexponential vectors.
\subsection{One-dimensional submodules} \label{SUBSECT:onedim}
Let~$M$ be a (left) $R$-module that is a finite-dimensional vector
space over~$F$. A submodule of~$M$ is said to be {\em
one-dimensional} if it is also a vector space of dimension one
over~$F$. Let~$N$ be a one-dimensional submodule of~$M$ and~$\vv$ a
non-zero element of~$N$. Then~$N$ is generated by~$\vv$ as a vector
space over~$F$. So we may write~$N=F\vv$. Moreover, for all~$r \in
R$, there exists~$f \in F$ such that~$r\vv=f\vv$.

We review some results concerning one-dimensional submodules, which
will help us describe one-dimensional submodules of a
finite-dimensional module over a Laurent-Ore algebra by a finite
amount of information, as sketched in~\cite[page 111]{VdpSinger03}
and~\cite{GrSch05} for differential modules.
\begin{lemma} \label{LM:sub}
Let $N_1, \ldots, N_s$ be one-dimensional submodules of {an
$R$-module} such that the sum~$\sum_{i=1}^s N_i$ is direct. If~$N$
is a nontrivial submodule contained in~$\sum_{i=1}^s N_i$, then
there exists a one-dimensional submodule~$N^\prime \subset N$.
Moreover,~$N^\prime$ is isomorphic to some~$N_i$.
\end{lemma}
\begin{proof}
Every element of~$\sum_{i=1}^s N_i$ can be (uniquely)
expressed as a sum of  elements in~$N_1$, \ldots, $N_s$.
Among all non-zero elements of~$N$, choose a~$\vv \in N$ such that
its additive expression is shortest. Without loss of generality,
suppose that the additive expression of~$\vv$ is~$\sum_{i=1}^t
\vv_i$ where $\vv_i\in N_i$ is nonzero and~$1 \le t \le s$. For any
$r \in R$, $r \vv_i \in N_i$, and, hence, $r \vv_i = a_i \vv_i $ for
some $a_i \in F$, {because $N_i$ is one-dimensional.} It follows
that $r \vv = \sum_{i=1}^t a_i \vv_i$. By the selection of~$\vv$, $r
\vv {-} a_1 \vv= 0$, and, hence, $F \vv$ is a one-dimensional
submodule in~$N$. Let~$\pi_1$ be the projection
from~$\bigoplus_{i=1}^s N_i$ to~$N_1$. Note that~$\pi_1(\vv)
{=}\vv_1 {\neq} 0$. So the restriction of~$\pi_1$ on~$F\vv$ is
bijective since~$F\vv$ and~$N_1$ both have
dimension~one.~\end{proof}

As a consequence, one can prove by induction on $s$ that
\begin{cor} \label{COR:direct}
{If~$N_1, \ldots, N_s$ are pairwise nonisomorphic one-dimensional
submodules of an $R$-module,} then~${\sum_{i=1}^s N_i}$ is direct.
\end{cor}

\smallskip
Let~$\cM_1$ be the set of all one-dimensional submodules of a
finite-dimensional $R$-module~$M$, and
$\overline{\cM}_1$ the set of equivalence classes of~$\cM_1$ modulo
isomorphism. The cardinality of~$\overline{\cM}_1$ is finite by
Corollary~\ref{COR:direct}. For an equivalence class~$I$
in~$\overline{\cM}_1$, there exist a finite number of
submodules~$N_1=F\vv_1$, \ldots, $N_s=F\vv_s$ in~$I$ such
that~$\vv_1$, \ldots $\vv_s$ are linearly independent over~$F$, and
moreover, for every $F\vv \in I$, $\vv$ is linearly dependent
on~$\vv_1$, \ldots $\vv_s$ over~$F$. Then~$\sum_{N \in I}N =
\bigoplus_{i=1}^s N_i$. Setting the latter (direct) sum to be~$S_I$,
one can prove, using Lemma~\ref{LM:sub} and induction, the following

\begin{prop} \label{PROP:sum}
With the notation just introduced, we have $ \sum_{N \in
\cM_1} N = \bigoplus_{I \in \overline{\cM}_1} S_{I}. $ \end{prop}
\subsection{{Hyperexponential vectors}} \label{SUBSECT:ext}
Let~$R$ be a ring and~$\Delta$ be a finite set of commuting maps
from~$R$ to itself. A map in~$\Delta$ is assumed to be either a
derivation on~$R$ or an automorphism of~$R$. Recall that a
derivation~$\delta$ is an additive map satisfying  the
multiplicative rule $\delta(ab) = a \delta(b) + \delta(a) b$ for all
$a, b \in R.$
The pair~$(R, \Delta)$ is called a $\Delta$-ring.

For a derivation~$\delta \in \Delta$, an element~$c$ of~$R$ is
called a {\em constant} with respect to~$\delta$ if~$\delta(c)=0$.
For an automorphism~$\sigma \in \Delta$,~$c$ is called a {\em
constant} with respect to~$\sigma$ if~$\sigma(c)=c$. An element~$c$
of~$R$ is called a {\em constant} if it is a constant with respect
to all maps~in~$\Delta$. The set of constants of~$R$, denoted
by~$C_R$, is a subring. The ring~$C_R$ is a subfield if $R$ is a
field. 

Let~$(F, \Delta)$ be a~$\Delta$-field and~$R$ a commutative ring
containing $F$. If all the maps in~$\Delta$ can be extended to~$R$
in such a way that all derivations (resp.\ automorphisms) of~$F$
become derivations (resp.\ automorphisms) of~$R$
and the extended maps commute pairwise, then~$(R, \Delta)$, or
simply~$R$, is called a $\Delta$-extension of~$F$.

In a $\Delta$-extension~$R$ of~$F$, a non-zero element~$h$ is said
to be~{\em hyperexponential with respect to a map $\phi$
in~$\Delta$} if $\phi(h){=}rh$ for some~$r \in F$. The element~$r$
is denoted~$\lphi(h)$. The element~$h$ is said to be {\em
hyperexponential} over~$F$ if it is hyperexponential with respect to
all the maps in~$\Delta$. A non-zero vector~$V \in R^n$ is said to
be {\em hyperexponential (with respect to a map~$\phi$)} if there
exist $h \in R$, hyperexponential (with respect to~$\phi$), and $W
\in F^n$ such that $V =hW$ {(see \cite[Chapter~4]{Wu_thesis})}. {A
straightforward calculation shows that}
\begin{lemma} \label{LM:ratio}
Let~$h_1, h_2$ be two hyperexponential elements of a
$\Delta$-extension~$E$ of~$F$. If $\lphi(h_1)=\lphi(h_2)$ for
all~$\phi \in \Delta$ and~$h_2$ is invertible, then~$h_1/h_2$ is a
constant.
\end{lemma}

\smallskip
Let~$\Delta^\prime$ be a nonempty subset of~$\Delta$, and let~$E$
and~$E^\prime$ be~$\Delta$ and $\Delta^\prime$-extensions of~$F$,
respectively. {The $F$-algebra~$E \otimes_F E^\prime$ is a
$\Delta^\prime$-extension,} where $\delta(r \otimes r^\prime) =
\delta(r) \otimes r^\prime + r \otimes \delta(r^\prime)$ and $\s(r
\otimes r^\prime)=\s(r) \otimes \s(r^\prime)$ for all derivation
operators~$\delta$ and automorphisms~$\sigma$ in~$\Delta^\prime$.
The canonical maps~$E \longrightarrow E \otimes_F E^\prime$
and~$E^\prime \longrightarrow E \otimes_F E^\prime$ are injective
since~$E$ and~$E^\prime$ are $F$-algebras. Thus~$E \otimes_F
E^\prime$ can be regarded as a $\Delta^\prime$-extension that
contains both~$E$ and~$E^\prime$.

\section{Modules over Laurent-Ore algebras} \label{SECT:mlo}
In the sequel, we set~$\Delta=\{\delta_1, \dots, \delta_\ell,
\sigma_{\ell+1}, \dots, \sigma_m\}$ where $\delta_1, \ldots,
\delta_\ell$ are derivation operators on~$F$ and~$\sigma_{\ell+1},
\ldots, \sigma_m$ are automorphisms of~$F$.

The {\em Laurent-Ore algebra} over~$F$ is a noncommutative ring
$\LO=F[\pa_1,\dots,\pa_m,\pa_{\ell+1}^{-1},\dots, \pa_m^{-1}]$ whose
multiplication rules are~$\pa_s\pa_t=\pa_t\pa_s$,
$\pa_j\pa_j^{-1}=1,$ $\pa_ia = a\pa_i+ \delta_i(a),$  $\pa_j a=
\sigma_j(a)\pa_j,$ and $\pa_j^{-1}a=\sigma_j^{-1}(a)\pa_j^{-1}$,
where $1\le s < t\le m$, $1\le i{\le} \ell$, $\ell+1\le j\le m$,
and~$a\in F$. The algebra~$\LO$ can be constructed from an Ore
algebra over~$F$ (see~\cite{BronsteinLiWu05}). For any
finite-dimensional~$\LO$-module, its~$F$-bases may be computed  via
the Gr\"obner basis techniques in~\cite[Chapter~3]{Wu_thesis}.

Let~$\Delta^\prime$ be a nonempty subset of~$\Delta$. Then~$\Delta^\prime$ corresponds
to a Laurent-Ore algebra~$\LO^\prime$. An $\LO$-module~$M$ is also an $\LO^\prime$-module.
To distinguish the different module structures, we write~$(M, \Delta)$ and~$(M, \Delta^\prime)$
to mean that~$M$ is an $\LO$-module and an $\LO^\prime$-module, respectively.

Let~$M$ be an $\LO$-module with a finite basis~$\vb_1, \ldots,
\vb_n$ over~$F$. The module structure of~$M$ is determined by~$m$
matrices $A_1, \ldots, A_m$ in~$F^{n \times n}$ such that
\begin{equation} \label{EQ:mstruct}
{\pa_i(\vb_1, \ldots, \vb_n)^{T}{=}A_i (\vb_1, \ldots, \vb_n)^T,
\,\, i{=}1, \ldots, m.}
\end{equation}
Note that $A_{\ell+1}$, \ldots, $A_m$ are invertible because~$\LO$
contains $\pa_{\ell+1}^{-1},$ \ldots, $\pa_m^{-1}$. We call $A_1$,
\ldots, $A_m$ {\em the structure matrices} with respect to~$\vb_1$,
\ldots, $\vb_n$. {For a vector~$Z=(z_1, \ldots, z_n)^T$ of
unknowns},
\begin{equation} \label{EQ:fint}
{\delta_i(Z){=}-A_i^T Z, \,\, i\le \ell, \quad
\sigma_j(Z){=}\left(A_j^{-1}\right)^T Z, \,\, j >\ell,}
\end{equation}
is called the system associated to~$M$ and the basis~$\vb_1$,
\ldots, $\vb_n$. Systems associated to different bases are
equivalent in the sense that the solutions of one system can be
transformed to those of another by a matrix in~$F^{n \times n}$.
{The commutativity of the maps in~$\Delta$ implies
that~(\ref{EQ:fint}) is fully integrable~\cite[Definition
2]{BronsteinLiWu05}.} A detailed verification of this assertion is
presented in~\cite[Lemma 4.1.1]{Wu_thesis}. On the other hand, every
fully integrable system is associated to its module of formal
solutions~\cite[Example 4]{BronsteinLiWu05}, which is
an~$\LO$-module of finite dimension.

A solution~$V$ of~(\ref{EQ:fint}) is called a hyperexponential
solution if~$V$ is a hyperexponential vector. It is called a
rational solution if {the entries of~$V$ are in~$F$.}

The next proposition connects one-dimensional submodules with
hyperexponential vectors.
\begin{prop}\label{PROP:corresp}
Let an~$\LO$-module~$M$ have a finite $F$-basis~$\vb_1,\dots, \vb_n$
with structure matrices given in~(\ref{EQ:mstruct}) and the
associated system in~(\ref{EQ:fint}). Let~$\vu=\sum_{i=1}^nu_i\vb_i$
with $u_i \in F$ not all zero.
\begin{enumerate}
\item[(i)] If there exists a hyperexponential element~$h$ in some
$\Delta$-extension  such that~$h(u_1, \ldots, u_n)^T$ is a solution
of~(\ref{EQ:fint}), then~$F\vu$ is a submodule of~$M$ with
\begin{equation}\label{EQ:rel3}
\mskip-29mu\pa_i(\vu){=}-\ld_i(h)\vu,  \, i \le \ell \,\, {\rm and}
\,\, \pa_j(\vu){=}\ls_j(h)^{-1}\vu,\, j >\ell.
\end{equation}
\item[(ii)] If~$F\vu$ is a submodule of~$M$ then there exists an
invertible hyperexponential element~$h$ in some $\Delta$-extension
 such that~$h(u_1,\dots, u_n)^T$  is a solution of~(\ref{EQ:fint}).
\end{enumerate}
\end{prop}
\begin{proof}
 Let~$U=(u_1, \ldots, u_n)^T$ and~$\vb =(\vb_1, \ldots,
\vb_n)^T$. If~$hU$ is a solution of~(\ref{EQ:fint}), then
$\delta_i(U) = - A_i^T U-\ld_i(h)U$ for~$i \le \ell$.
Therefore $ \pa_i (\vu) 
=\delta_i \left(U^T \right)\vb+U^T A_i\vb
=-\ld_i(h)\vu $ for~$i \le \ell$.
Similarly,~$\pa_j(\vu)=\ls_j(h)^{-1}\vu$ for~$j> \ell$. So~$F\vu$ is
a submodule and~\eqref{EQ:rel3} holds.

Now let~$F\vu$ be a submodule. Then~$\pa_i \vu = f_i \vu$ where~$f_i
{\in} F$ for~$1\le i\le m,$ and~$f_j \neq 0$ for~$j>\ell$. The
system
associated to~$F\vu$ 
is $\{\, \delta_i(z)=-f_iz,   i\le \ell,\, \sigma_j(z)=f_j^{-1}z, j>
\ell \}.$
By Theorem~1 in~\cite{BronsteinLiWu05} it has an invertible
solution~$h$ in certain~$\Delta$-extension. Thus~$h$ is
hyperexponential over~$F$. From $\pa_i(\vu)= f_i\vu$ it follows
that~$\delta_i(U)=f_iU-A_i^TU$, which together with
$\ld_i(h)=-f_i$ implies~$\delta_i(hU)=-A_i^ThU$ for~$i\le \ell$.
Similarly, we get~$\sigma_j(hU)=(A_j^{-1})^ThU$
for~$j>\ell$.~\end{proof}

Let $h_1$ and~$h_2$ be two hyperexponential elements of a
$\Delta$-extension of~$F$ such that~$h_1(u_1, \ldots, u_n)^T$
and~$h_2(v_1, \ldots, v_n)^T$ are solutions of~(\ref{EQ:fint}). From
Proposition~5~(i),~$F\vu$ and~$F\vv$ with~$\vu=\sum_{i=1}^nu_i
\vb_i$ and $\vv=\sum_{i=1}^n v_i \vb_i$ are  one-dimensional
submodules of~$M$. Suppose~$F\vu = F\vv$. Then~$\vu = r\vv$ for
some~$r \in F$, {which, together with~(\ref{EQ:rel3}), implies that
$\lphi(rh_1) = \lphi(h_2)$} for all $\phi \in \Delta$. By
Lemma~\ref{LM:ratio},~$rh_1=c\,h_2$ with~$c$ a constant if we assume
that~$h_2$ is invertible. Consequently, $h_1(u_1,
\ldots,u_n)^T=c\,h_2(v_1, \ldots, v_n)^T$. In the situation
described in Proposition~\ref{PROP:corresp}, we say that the
hyperexponential vector~$h(u_1, \ldots, u_n)^T$ corresponds to the
submodule~$F\vu$ and understand that in any $\Delta$-extension, this
correspondence is unique up to constant multiples.

The next lemma tells us how to decide whether two one-dimensional submodules are isomorphic.
\begin{lemma} \label{LM:iso1}
Let~$M$ be an $\LO$-module with a finite $F$-basis $\vb_1, \ldots,
\vb_n$. Let $\vu=\sum_{i=1}^n u_i \vb_i$ and $\vv=\sum_{i=1}^n v_i
\vb_i$ where~$u_i$, $v_j \in F$. Suppose that~$F\vu$ and~$F\vv$ are
two one-dimensional submodules of~$M$ and that~$\pa_i \vu = f_i \vu$
and $\pa_i \vv = g_i \vv$, where $f_i, g_i \in F$ and $i=1, \ldots,
m$. Then we have the following statements:
\begin{enumerate}
\item[(i)] The  map $\vu \mapsto r \vv$ from $F\vu$ to $F\vv$ is a module
isomorphism  if and only if $r$ is a nonzero solution of the system
\begin{equation} \label{EQ:iso1}
\delta_i(z)=(f_i-g_i)z, \,\, i\le \ell, \,\,\, \sigma_j(z)= f_j
g_j^{-1}\,z, \,\, j>\ell.
\end{equation}
\item[(ii)] Suppose that $h(u_1, \ldots, u_n)^T$ is a solution of the
system
associated to~$M$, where~$h$ is hyperexponential 
in some $\Delta$-extension of~$F$. Then~$F\vu$ and~$F\vv$ are
isomorphic if and only if there exists a non-zero $r \in F$ such
that $rh(v_1, \ldots, v_n)^T$ is a solution of the associated
system.
\end{enumerate}
\end{lemma}
\begin{proof}
 Let $\psi:F\vu \rightarrow F\vv$ be a module
isomorphism~with $\psi(\vu)=r\vv$ for some non-zero~$r \in F$.
It follows that
\[ \psi(\pa_i\vu)=f_i r \vv = \pa_i (r \vv) =
\left\{ \begin{array}{ll}
(\delta_i(r) + g_i r) \vv, &   i \le \ell, \smallskip \\
\sigma_i(r)g_i\vv, &  i > \ell.
\end{array} \right.
\]
Thus~$r$ is a non-zero solution of~(\ref{EQ:iso1}). Conversely,
if~$r$ is a non-zero solution of~(\ref{EQ:iso1}), then $\vu \mapsto
r\vv$ gives rise to a module isomorphism from~$F\vu$ to~$F\vv$ by a
similar calculation.

To prove (ii), we assume that the module structure of~$M$ is given
by~(\ref{EQ:mstruct}) and the associated system is given
by~(\ref{EQ:fint}). Thus $f_i=-\ld_i(h)$ and $f_j = \ls_j(h)^{-1}$
by Proposition~\ref{PROP:corresp} (i).

If~$F\vu\rightarrow F\vv$ is an isomorphism given by~$\vu\mapsto
r\vv$ with~$r{\in} F$, then $r$ satisfies (\ref{EQ:iso1}) by (i),
hence
\begin{equation} \label{EQ:rel4}
g_i = -\ld_i(rh), \,\, i \le \ell, \,\,\,{\rm and} \,\,\, g_i =
\ls_i(rh)^{-1}, \,\, i>\ell.
\end{equation}
Set $V=(v_1,\dots, v_n)^T$. From~$\pa_i(\vv)= g_i\vv$, we get
$\delta_i(V)= g_iV-A_i^TV$ for~$i\le \ell$, which together
with~\eqref{EQ:rel4} implies
$$
\delta_i(r\,hV)   = \delta_i(rh)V + r\,h\,\delta_i(V)
=-A_i^Tr\,hV.
$$
A similar calculation yields $\sigma_j(rhV){=} (A_j^{-1})^TrhV$
for~$j>\ell$. So~$rhV$ is a solution of~(\ref{EQ:fint}).

Conversely, let 
$rh(v_1, \ldots, v_n)^T$ with $r\in F$ be a solution
of~(\ref{EQ:fint}). From Proposition~\ref{PROP:corresp} (i),
both~$F{\bf u}$ and~$Fr{\bf v}$ $(=F\vv)$ are two submodules, and in
addition, $ \pa_i(\vu)=-\ld_i(h)\vu$ and~$
\pa_i(r\vv)=-\ld_i(h)r\vv$ for~$i \le \ell$, and
$\pa_j(\vu)=\ls_j(h)^{-1}\vu$ and~$\pa_j(r\vv)=\ls_j(h)^{-1}r\vv$
for~$j> \ell$. One can then verify easily that~$\vu \mapsto r\vv$ is
an isomorphism.
\end{proof}

We now construct a $\Delta$-extension~$E$ of~$F$ such that every
one-dimensional submodule of~$M$ corresponds to a hyperexponential
vector~$hV$, where~$h$ is an invertible element of~$E$ and~$V$ is a
column vector in~$F^n$. Denote by~$\cM_1$  the set of
one-dimensional submodules of~$M$ and by~$\overline{\cM}_1 =\{I_1,
\ldots, I_s\}$ the set $\cM_1$ modulo isomorphism. {For each~$k$
in~$\{1, \ldots, s\}$, we select a one-dimensional submodule~$N_k$
in~$I_k$. Assume that~$N_k$ corresponds to a hyperexponential
vector~$h_kV_k$, where $h_k$ is in some~$\Delta$-extension of~$F$
and~$V_k$ is a vector with entries in~$F$. We can verify directly
that the system} $$ \left\{\begin{array}{l l}
 \delta_i(Z)= {\rm diag}(\,\ld_i(h_1),\,
\ldots,\, \ld_i(h_s)\,)Z, & 1\le i \le \ell,\medskip \\
 \sigma_j(Z)= {\rm
diag}(\ls_j(h_1), \ldots, \ls_j(h_s))Z,  & \ell < j\le m
\end{array}\right.
$$
where $Z=(z_1, \ldots, z_s)^T$, is fully integrable.
By 
{Theorem~1 in~\cite{BronsteinLiWu05}, there exists
a~$\Delta$-extension~$E$ containing a fundamental matrix~${\rm
diag}(h_1^\prime, \dots, h_s^\prime)$
 and the
inverse of its determinant.} {That is, for every~$k$ with~$1 \le k
\le s$,~$\lphi(h_k)=\lphi(h^\prime_k)$ for all~$\phi \in \Delta$.
Consequently, $h_k^\prime V_k$ also corresponds to~$N_k$. By
Lemma~\ref{LM:iso1} (ii), we need only to search for
hyperexponential solutions of the system associated to~$M$ in~$E^n$
to determine~$\cM_1$.} Observe that the construction of~$E$ is
independent of the choices of $F$-bases, since all associated
systems are equivalent. The ring~$E$ is therefore called a {\em
hyperexponential extension} relative to~$(M, \Delta)$.

Next, we represent all one-dimensional submodules of~$M$ by a finite
amount of information. Suppose that~$M$ has a basis~$\vb_1$, \ldots,
$\vb_n$ and the associated system~(\ref{EQ:fint}), with~$\cM_1$
and~$\overline{\cM}_1 =\{I_1, \ldots, I_s\}$ denoted as above.
Let~$E$ be a hyperexponential extension of~$M$. {By
Lemma~\ref{LM:iso1} (ii), for each~$k$ with~$1 \le k \le s$ there
exists an invertible hyperexponential element~$h_k$ of~$E$ such that
every $N \in I_k$ corresponds to a solution~$h_kV_{k,N}$
of~(\ref{EQ:fint}), where~$V_{k,N}$ is a column vector in~$F^n$.}
Let~$V_k$ be a matrix whose column vectors form a maximal set of
$F$-linearly independent vectors among all~$V_{k,N}$ for~$N \in
I_k$. We call $\{(h_1, V_1), \ldots, (h_s, V_s)\}$ a {\em
representation of hyperexponential solutions} of~(\ref{EQ:fint})
and~$\{V_1, \ldots, V_s\}$ a {\em representation of~$\cM_1$}
relative to the given basis.
\begin{prop} \label{PROP:hsol}
With the notation just introduced, let~$\bar{E}$ be a
$\Delta$-extension containing~$E$, and~$H$ the set of
hyperexponential solutions of~(\ref{EQ:fint}) in~$\bar{E}$. If
$\{(h_1, V_1), \ldots, (h_s, V_s)\}$ is a representation of
hyperexponential solutions of~(\ref{EQ:fint}), then
\begin{itemize}
\item[(i)] $N$ is a one-dimensional submodule of~$M$ if and only
if~$N$ is generated by $U^T(\vb_1, \ldots, \vb_n)^T$ over~$F$,
where~$U$ is a $C_F$-linear combination of the column vectors in
some~$V_k$;
\item[(ii)] $H$ is the disjoint union $\cup_{k=1}^s H_k$, where
$H_k {=} \left\{ c \, h_k V_k C \right\}$ with~$c$  an arbitrary
non-zero constant in~$\bar{E}$ and~$C$  an arbitrary non-zero column
vector over~$C_F$.
\end{itemize}
\end{prop}
\begin{proof}
Observe that for every~$k \in \{1, \ldots, s\}$ and~$N {\in} I_k$,
the vector~$V_{k,N}$ defined above is a rational solution of the
system obtained by substituting $h_k(z_1, \ldots, z_n)^T$
into~(\ref{EQ:fint}). Hence by Lemma~1.7 in~\cite{VdpSinger03} and
its difference analogue, the column vectors in~$V_k$ also form a
maximal set of $C_F$-linearly independent {vectors} among all~$V_{k,
N}$ for~$N \in I_k$. Consequently,~$V_{k,N}$ is a $C_F$-linear
combination of the column vectors in~$V_k$, which proves the first
assertion.

Clearly, $\cup_{k=1}^s H_k \subset H$ and $H_i \cap H_j = \emptyset$
for all~$i \neq j$. Assume~$hW\in H$ with $h \in \bar{E}$ and~$W$ a
column vector in~$F^n$. By Proposition~\ref{PROP:corresp} (i), $hW$
corresponds to a one-dimensional submodule~$N$, which  by~(i) also
corresponds to a hyperexponential solution~$h_k V_{k, N} \in H_k$
for some~$k$ with~$1 \le k \le s$. Thus the two solutions differ
from a constant multiple according to the discussion following the
proof of Proposition~\ref{PROP:corresp}.
\end{proof}

\section{Parametric hyperexponential \\ vectors} \label{SECT:hsol}
As before, let $\Delta=\{\delta_1, \ldots, \delta_\ell,
\sigma_{\ell+1}, \ldots, \sigma_m\}$ where the $\delta_i$ and
$\sigma_j$ are derivation operators and  automorphisms of~$F$,
respectively, and~$\LO$ be the Laurent-Ore algebra over~$F$.
Let~$M$ be an $n$-dimensional $\LO$-module with an associated
system given in~(\ref{EQ:fint}). For the purpose of this article,
it suffices to find the hyperexponential solutions
of~(\ref{EQ:fint}) in a hyperexponential extension relative
to~$M$. We plan to proceed as follows:

\smallskip \noindent
First, compute hyperexponential solutions of a matrix equation
in~(\ref{EQ:fint}), say, $\delta_1(Z)=-A_1^T Z$. The set of
solutions is partitioned into finitely many groups by
Proposition~\ref{PROP:hsol}. Each group is given as~$
c \,
h \, V \, C$, where~$c$ is a constant with respect
to~$\delta_1$,~$h$ is hyperexponential with respect to~$\delta_1$,
$V$ is a matrix over~$F$, and $C$ is a column vector whose entries
 are arbitrary constants with respect to~$\delta_1$.

\smallskip \noindent
Second, substitute~$Z{=}\, c \, h  V \, C$ into another matrix
equation, say~$\delta_2(Z)=-A_2^T Z$ to find~$c$ and~$C$ so that~$Z$
is a hyperexponential solution of {both~$\delta_1(Z)=-A_1^T Z$
and~$\delta_2(Z)=-A_2^T Z$. There arise several questions in this
process:}
\begin{itemize}
\item[(a)] In what extensions do we compute hyperexponential
solutions of~$\delta_1(Z){=}-A_1^T Z$ ? \item[(b)] Does the
substitution introduce coefficients outside~$F$~? Note that~$h$ is
not necessarily  hyperexponential with respect to~$\delta_2$.
\item[(c)] {How do we determine~$c$ and $C$ so that~$Z$ is
hyperexponential with
           respect to both~$\delta_1$ and~$\delta_2$?}
\end{itemize}
These questions will be answered in Proposition~\ref{PROP:ext} at the end of this section.

Let~$\Theta$ be the commutative monoid generated by the $\delta_i$
and~$\sigma_j$. The multiplication in~$\Theta$ is the composition of
maps. For~$\theta=\delta_1^{k_1}\ldots\delta_{\ell}^{k_{\ell}}
\sigma_{\ell+1}^{k_{\ell+1}}\ldots\sigma_{m}^{k_m}\in \Theta,$ the
sum $\sum_{i=1}^{m}k_i$ is called the {\em order}~of~$\theta$. The
set of elements of~$\Theta$ of order less than or equal to~$s$ is
denoted~$\Theta_s$.
\begin{lemma} \label{LM:wr-cas}
Let~$F$ be a $\Delta$-field,
$f_1, \ldots ,f_s$ in~$F$ and~$E$ a $\Delta$-ring extension of~$F$.
Then~$f_1, \ldots, f_s$ are linearly dependent over $C_E$ if and only if the matrix
$
W(f_1, \ldots, f_s)=\left(\theta f_i\right)_{\theta\in\Theta_{s-1}, 1\leq
i\leq s}
$
has rank less than~$s$.
In particular, if $f_1, \ldots ,f_s\in F$ are linearly dependent over~$C_E$,
then they are linearly dependent over~$C_F$.
\end{lemma}
\begin{proof}
If there exist~$d_1, \ldots, d_s \in C_E$, not all zero, such that
$d_1f_1+ \ldots + d_sf_s = 0$, then $d_1 \theta f_1 + \ldots + d_s
\theta f_s = 0$ for all $\theta \in \Theta_{s-1}$. The
matrix~$W(f_1, \ldots, f_s)$ has rank less than~$s$ by
Corollary~4.17  in~\cite[Chapter XIII \S 4]{Lang}.

{Assume that~$W(f_1, \ldots, f_s)$} has rank less than~$s$. The
proof follows the similar arguments concerning Wronskians and
Casoratians, and proceeds by induction on~$s$. The statement holds
when~$s=1$. Assume that~$s>1$ and that the statement holds for lower
values of~$s$. We can find in~$F$ a nontrivial solution $c_1, \ldots
, c_s$ to the equations $\sum_{k=1}^s c_k \theta (f_k) = 0$ for all
$\theta \in \Theta_{s-1}$. Since~$F$ is a field, we can assume $c_1
= 1$. Applying~$\delta_i$ (resp.~$\sigma_j$) to each equation
indexed by~$\theta \in \Theta_{s-2}$ and then subtracting from the
equation indexed by~$\delta_i\theta$ (resp.~$\sigma_j \theta$, and
noting that~$\sigma_j$ is an automorphism), we have~$ \sum_{k=2}^s
\delta_i(c_k) \theta (f_k) = 0$ and {$\sum_{k=2}^s (c_k -
\sigma_j^{-1}(c_k)) \theta (f_k) = 0$ for all~$\theta \in
\Theta_{s-2}$.} Either the~$c_k$ are constants or some
$\delta_i(c_k) \neq 0$ or some~$\sigma_j(c_k)-c_k \neq 0$. In the
former case, we have the conclusion. In the latter two cases, {the
matrix} $W(f_2, \ldots, f_s)$ has rank less than~$s-1$. The
induction hypothesis then implies that~$f_2, \ldots, f_s$ are
already linearly dependent over~$C_E$.
The conclusion of the
lemma is again satisfied.
\end{proof}

\begin{lemma} \label{LM:lsol}
Let~$K$ be a field and $R$ a commutative $K$-algebra. Let {
\[ \left\{
\begin{array}{ll}
 \sum_{j=1}^n a_{ij}X_j =0, & \,\, 1 \le i \le p \\ \\
 \sum_{j=1}^n b_{kj}X_j\ne 0,  & \,\, 1\le k\le q
      \end{array} \right.
\]}
be a system of equations with coefficients in~$K$. This system has a
non-zero solution in~$K$ if and only if it has a non-zero solution
in $R$.
\end{lemma}
\begin{proof}
 Let~$\{\alpha_j \}
$ be a $K$-basis of~$R$ and let $c_i = \sum_j d_{ij}\alpha_j$ with~$
d_{ij} \in K$ be a solution of the above system in~$R$. Substituting
in the system and equating the coefficients of the~$\alpha_i$, we
find a solution in~$K$.
\end{proof}

\noindent {\bf Notation:} In the rest of this article
$\Delta^\prime$ is a nonempty subset of~$\Delta$. For a
$\Delta$-ring~$R$, the ring of constants with respect to the maps
in~$\Delta^\prime$ is denoted~$C^\prime(R)$.

\begin{lemma} \label{LM:ext}
Let~$F$ be a $\Delta$-field,~$E$ a $\Delta$-extension of~$F$,
and~$E^\prime$ a $\Delta^\prime$-extension of~$E$. {Let $V_1, \ldots
, V_s, W$ be non-zero column vectors in~$F^n$, $c_1, \ldots ,c_s \in
C^\prime(E^\prime)$,  $g\in E$, $h\in E^\prime$ with $g, h$
invertible, and~$g W = h \sum_{i=1}^s c_i V_i$.} If~$h$ is
hyperexponential over $F$ with respect to~$\Delta^\prime$, and~$g$
is hyperexponential over $F$ with respect to~$\Delta$, then there
exist $d_1, \ldots , d_s \in C^\prime(F)$, $\bar{h} \in E$
with~$\bar{h}$ invertible such that the following statements hold:
\begin{itemize}
\item[(i)] $\bar{h}$ is hyperexponential over~$F$ with respect to~$\Delta$.
\item[(ii)] $\lphi (\bar{h}) = \lphi(h)$ for all~$\phi \in \Delta^\prime$.
\item[(iii)]$g W =  \bar{h} \sum_{i=1}^s d_iV_i$.
\end{itemize}
\end{lemma}
\begin{proof}
Let $W =(w_1, \ldots ,w_n)^T$ and~$V_i = (v_{1i}, \ldots,
v_{ni})^T$. Assume that~$w_1, \ldots , w_t$ are non-zero
while~$w_{t+1}, \dots, w_n$ are all zero. The equation $gW = h\sum_i
c_i V_i$ translates to
\begin{equation} gh^{-1}  =
\sum_ic_i\frac{v_{ji}}{w_j}, \quad\mbox{for } j=1,\ldots , t,
\label{eqn1}
\end{equation}
and~$0  =  \sum_ic_iv_{ki}$ for~$k=t+1,\ldots , n$. Note that the
equations~(\ref{eqn1}) imply that~$\sum_ic_i\frac{v_{ji}}{w_j} =
\sum_ic_i\frac{v_{li}}{w_l}$ for~$1\leq j, \, l \leq t.$
Furthermore, we have that for any $\phi \in \Delta^\prime$ there
is a $u_\phi \in F$ such that~$\lphi(\sum_ic_i\frac{v_{ji}}{w_j})
= \lphi(gh^{-1}) = u_\phi$ for~$1\leq j \leq t.$ Consider the
equations \begin{eqnarray}
\sum_ic_i\frac{v_{ji}}{w_j} &=& \sum_ic_i\frac{v_{li}}{w_l} \neq 0, \ \ 1\leq j,\,l \leq t \label{eqn3}\\
0 & = & \sum_ic_iv_{ki}, \ \  k=t+1,\ldots , n \label{eqn4}\\
\sum_ic_i\phi\left(\frac{v_{ji}}{w_j}\right) & = &
\sum_ic_i\frac{v_{ji}}{w_j}u_\phi, \ \ 1\leq j\leq t, \phi \in
\Delta^\prime. \label{eqn5}
\end{eqnarray}
Letting $\{\alpha_s\}$ be a $C^\prime(F)$-basis of $F$,  there exist
$a_s^{ji}$, $b_s^{ki}$, $c_s^{ji\phi}$,~$d_s^{ji\phi}$ in
$C^\prime(F)$ such that~$ \frac{v_{ji}}{w_j}  {=}  \sum_s a_s^{ji}
\alpha_s$,~$v_{ki} {=} \sum_s b_s^{ki} \alpha_s$, $
\phi\left(\frac{v_{ji}}{w_j}\right)  = \sum_s c_s^{ji\phi} \alpha_s$
and~$\frac{v_{ji}}{w_j}u_\phi  = \sum_s d_s^{ji\phi} \alpha_s$.
Substitute these into equations~(\ref{eqn3}), (\ref{eqn4}) and
(\ref{eqn5}). Using~Lemma~\ref{LM:wr-cas} and equating coefficients
of the~$\alpha_s$, we see that $X_i = c_i$ {satisfy} the following
system of equations for all~$s$:
\begin{eqnarray*}
\sum_i X_ia_s^{ji} & = & \sum_i X_ia_s^{li}, \quad  1\leq j,\,l
\leq
t \\
0 & = &\sum_iX_i b_s^{ki}, \quad   k=t+1,\ldots , n \\
\sum_iX_i c_s^{ji\phi}  & = &  \sum_iX_i d_s^{ji\phi}, \quad 1\leq
j\leq t, \, \phi \in \Delta^\prime,
\end{eqnarray*}
and that for~$1 \leq j \leq t$ there is an~$s$ such that $\sum_i
X_ia_s^{ji} {\neq} 0$. Lemma~\ref{LM:lsol} implies that this system
will have a solution $X_i {=} d_i$ in~$C^\prime(F)$. Let~$S {=}
\sum_i d_i\frac{v_{1i}}{w_1}{=} \ldots {=}\sum_i
d_i\frac{v_{ti}}{w_t} \neq 0$. Note that $\phi(S) = u_\phi S$ for
all $\phi \in \Delta^\prime$.  Therefore
$\lphi\left(\frac{g}{hS}\right) = 0$ for~all derivations $\phi \in
\Delta^\prime$ and $ \lphi\left(\frac{g}{hS}\right) = 1$ for all
automorphisms $\phi \in \Delta^\prime $ and so $g = Shd$ for some
$d\in C^\prime(E^\prime)$, that is, $w_jg = h(\sum_id_iv_{ji})d$ for
all $1\leq j \leq t$. {Letting $\overline{h} = hd =
w_jg/(\sum_id_iv_{ji})\in E$}, we have that $\ell\phi(\overline{h})
=\lphi(h)$ for $\phi \in \Delta^\prime$ and $\ell\phi(\overline{h})
\in F$ for all $\phi \in \Delta$.
\end{proof}

{We now consider how to have some information about~$\bar{h}$ given
in the conclusion of Lemma~\ref{LM:ext} without knowing~$gW$.
Let~$\Delta^\prime{=}\{\delta_1, \ldots, \delta_p, \,
\sigma_{\ell+1}, \ldots, \sigma_q\}$, $r_i{=}\ld_i(h)$ and $r_j{=}
\ls_j(h)$ where~$1 \le i \le p$ and~$\ell+1 \le j \le q$.} Note that
$r_i=\ld_i(\bar{h})$ and~$r_j=\ls_j(\bar{h})$ by Lemma~\ref{LM:ext}.
Assume that~$\phi \in \Delta \setminus \Delta^\prime$. We want to
compute an element~$r$ of~$F$ such that~$r=\lphi(\bar{h})$.

\medskip \noindent
{\bf Case 1.} $\phi$ is a derivation operator. On one hand, we
have~$\phi \circ \delta_i(\bar{h})= (\phi(r_i)+r_ir) \bar{h}$ and
$\phi \circ \s_j(\bar{h}) = (\phi(r_j)+r_jr) \bar{h}$. On the other
hand, we have $ \delta_i \circ \phi (\bar{h})= (\delta_i(r)+r_ir)
\bar{h}$ and $\s_j \circ \phi (\bar{h}) = \s_j(r) r_j \bar{h}.$ By
the commutativity of the maps in~$\Delta$, $r$ is a solution of the
system {\begin{equation} \label{EQ:der} \left\{
\begin{array}{ll}
\delta_i(z) = \phi(r_i), & \,\,  1 \le i \le  p \\ \\
\sigma_j(z) - z = \lphi(r_j), & \,\, \ell + 1 \le j \le q.
\end{array} \right.
\end{equation}
Consequently, if there exists~$h$ such that~$gW{=}h\sum_i c_i V_i$, (\ref{EQ:der}) has a solution~$r$ in~$F$
and~$\lphi(\bar{h}) {=} r{+}c$ for some $c \in C^\prime(F)$.}

\medskip \noindent
{\bf Case 2.} $\phi$ is an automorphism. A similar calculation shows
that~$r$ is a non-zero solution of the system {\begin{equation}
\label{EQ:auto} \left\{ \begin{array}{ll}
\delta_i(z) = (\phi(r_i){-}r_i) z, & \,\,  1 \le i \le  p \\ \\
{\sigma_j}(z) = \lphi(r_j) z,  & \,\, \ell + 1 \le j \le q.
\end{array} \right.
\end{equation}
Consequently, if there is~$h$ such that~$gW=h\sum_i c_i V_i$, (\ref{EQ:auto}) has a solution~$r$ in~$F$
and~$\lphi(\bar{h})=cr$ for some $c \in C^\prime(F)$.}

\smallskip
Let~$h$ be a hyperexponential element with respect
to~$\Delta^\prime$. We say that~$h$ is {\em extensible} for a
map~$\phi \in \Delta \setminus \Delta^\prime$ if there
exists~$\bar{h}$, hyperexponential with respect to
both~$\Delta^\prime$ and~$\phi$, such
that~$\lpsi(h)=\lpsi(\bar{h})$ for all~$\psi \in \Delta^\prime$.

{By the above discussion,~$h$ is extensible for a derivation (resp.\
an automorphism) if and only if (\ref{EQ:der}) (resp.\
(\ref{EQ:auto})) has a rational solution.}

\begin{prop} \label{PROP:ext}
Let~$\Delta \setminus \Delta^\prime$ have one element and~$M$ be an
$\LO$-module of finite dimension. Let~$E$ and~$E^\prime$ be
hyperexponential extensions relative to~$(M, \Delta)$ and~$(M,
\Delta^\prime)$, respectively. Let~$\cA$ and~$\cA^\prime$ be the
systems associated to~$(M, \Delta)$ and~$(M, \Delta^\prime)$,
respectively. Let~$\{(h_1^\prime, V_1^\prime), \ldots, (h_t^\prime,
V_t^\prime)\}$ be a representation of hyperexponential solutions
of~$\cA^\prime$ in~$E^\prime$ with respect to~$\Delta^\prime$. Then
there exist a $\Delta^\prime$-extension~$R$ of~$F$ containing
both~$E$ and~$E^\prime$, and 
invertible hyperexponential elements~$h_1, \ldots, h_s$ in $R$, with
$s\le t$, such that, for every hyperexponential solution~$gW$
of~$\cA$ with coordinates in~$R$, $g W {=} h_k V_k^\prime D$, where~$k$ is unique
and~$D$ is a hyperexponential vector over~$C^\prime(R)$.
\end{prop}
\begin{proof}
Let~$\Delta \setminus \Delta^\prime {=}\{\phi\}$. Assume that
$h_1^\prime, \ldots, h_s^\prime$ are extensible to~$h_1, \ldots,
h_s$ for~$\phi$, respectively, while~$h_{s+1}^\prime,$ \ldots,
$h_t^\prime$ are not extensible. We can regard~$h_1$, \ldots, $h_s$
as invertible elements in a~$\Delta$-extension~$E^{\prime\prime}$,
as we did in the construction of hyperexponential extensions.
Let~$R=E \otimes_F E^\prime \otimes_F E^{\prime\prime}$. Since~$gW$
is a hyperexponential solution of~$\cA$, it is a hyperexponential
solution of~$\cA^\prime$. By Proposition~\ref{PROP:hsol}, {there
exist~$k$ with~$1 \le k \le t$} and a column vector~$C$ with entries
in~$C^\prime(R)$ such that $gW=h_k^\prime V_k^\prime C$. By
Lemma~\ref{LM:ext} we have $gW=\bar{h}V_k^\prime D^\prime$ where
$\bar{h} \in R$ is hyperexponential such that
$\lpsi(h_k^\prime)=\lpsi(\bar{h})$ for all~$\psi \in \Delta^\prime$,
and~$D^\prime$ is a column vector with entries in~$C^\prime(F)$.
Hence,~$h_k^\prime$ is extensible and the ratio~$d = \bar{h}/h_k$ is
in~$C^\prime(R)$ by Lemma~\ref{LM:ratio}. Setting~$D=dD^\prime$
yields the proposition.
\end{proof}
\section{Algorithm description}~\label{SECT:alg}
Let~$M$ be an $\LO$-module with an $F$-basis $\vb_1$, \ldots,
$\vb_n$ and let~$\vb=(\vb_1, \ldots, \vb_n)^T$. We will compute
one-dimensional submodules of~$M$ recursively. The key step for
recursion proceeds as follows.

Assume that we have obtained all one-dimensional submodules of~$(M,
\Delta^\prime)$, where~$|\Delta \setminus \Delta^\prime|=1$.
Let~$\cA$ and~$\cA^\prime$ be the systems associated to~$(M,
\Delta)$ and~$(M, \Delta^\prime)$, respectively. Let~$U$ be an $n
\times s$ matrix over~$F$ such that the set
$$
\begin{array}{rl}
 S= \left\{\mskip-15mu\right. &F\vu \mid \vu= (U
\,C)^T  \vb  \mbox{ and }\\
& \left.  C \mbox{ is a nonzero column vector over }
C^\prime(F)\right\}
\end{array}
$$
is an equivalence class of one-dimensional submodules of $(M,
\Delta^\prime)$ with respect to isomorphism. If~$W^T\vb$ with~$W
{\in} F^n$ generates a one-dimensional submodule of~$(M, \Delta)$
that is in~$S$, then
there exists an element~$g$ in a hyperexponential extension relative
to $(M, \Delta)$ such that $g W$ is a solution of~$\cA$. By
Proposition~\ref{PROP:ext} there exists a hyperexponential
element~$h$ in some $\Delta^\prime$-extension~$R$ such that $g W=h U
D$ for some hyperexponential vector~$D$ with entries
in~$C^\prime(R)$. Moreover,~$h$ can be found by computing rational
solutions of equations~(\ref{EQ:der}) or~(\ref{EQ:auto}).
Substituting~$h U D$ into the matrix equation corresponding to the
map~in~$\Delta \setminus \Delta^\prime$, we get an ordinary
differential or difference matrix equation in~$D$ over~$F$.
This system translates to a system over~$C^\prime(F)$ by the
technique used in the proof of Lemma~\ref{LM:ext}, since we only
look for hyperexponential solutions in~$C^\prime(R)$. In this way we
obtain all one-dimensional submodules of~$(M, \Delta)$ that are
in~$S$.

To make this idea effective, we will need several assumptions.
Define $\Delta_0 = \emptyset$, $\Delta_i =\{\phi_1,
 \ldots, \phi_i\}$ and
 $C_i$ to be the set of all elements of $F$ that are constants with
 respect to~$\Delta_i$.
 Note that $C_0 = F$, $C_m = C_F$ and that each $C_i$ is a $(\Delta\backslash \Delta_i)$-field.
 The above algorithm can be formalized if we assume that, for each $i$,
 \begin{enumerate}
 \item One is able to identify the field $C_i$ and effectively carry out computations in  $C_i$
 as a $(\Delta\backslash \Delta_i)$-field.  Furthermore, we assume that we can find a $C_i$-basis
 of $F$ and express any element in $F$ in this basis.
  \item Assuming that $\phi_{i+1}$ is a derivation, we can decide if systems of the
  form $\{L_j(z) = a_j\mid a_j \in F\}_{j=1}^i$  have solutions in $F$ where $L_j(z) = \phi_j(z)$ if $\phi_j$
  is a derivation and $L_j(z) = \phi_j(z)-z$ if $\phi_j$ is an automorphism,
 and, if so, find one.
 \item Assuming that $\phi_{i+1}$ is an automorphism, we can decide if systems of
 the form $\{\phi_j(z) = a_jz \ | \ a_j \in F\}_{j=1}^i$ have  solutions in $F$, and,
 if so, find one.
 \item Given an equation~$\phi_{i+1} (Z) = AZ$ with $A \in C_i^{n\times
 n}$,
 we can find all hyperexponential solutions over $C_i$.
 \end{enumerate}

By conditions~$2$ and~$3$, we can find rational solutions
of~\eqref{EQ:der} and~\eqref{EQ:auto}.
 In condition $4$, if $\phi_{i+1}$ is a differential
 operator, methods for solving such an equation or reducing the system to
 a scalar equation and solving the scalar equation for certain fields are
 discussed in \cite{Barkatou99differential,BarkPflueg99,Bronstein92linear,vanHoeij97,Singer91}. Methods to find
 hypergeometric solutions for scalar difference equations are discussed in \cite{AbrPaulePetk98, vanHoeij99, Petk92}.
 We will discuss below a method to reduce systems to scalar equations in the difference~case.

\subsection{Ordinary case} \label{SUBSECT:ordinary}
Let~$\phi$ be a difference operator. Consider a system
\begin{equation} \label{EQ:sys1}
\phi(Z)=AZ \quad \mbox{with $A \in F^{n \times n}$ and $Z=(z_1,
\ldots, z_n)^T$}.
\end{equation}
From~\eqref{EQ:sys1}, we construct by linear algebra a linear
difference equation with minimal order, say,
$$ L(z_1)=\phi^k(z_1) +
a_{k-1}\phi^{k-1}(z_1) + \cdots + a_0 z_1 = 0
$$
where~$a_i \in C^\prime(F)$. If~$k=n$, then each of the~$z_i$ is a
linear combination of~$z_1, \phi(z_1), \ldots, \phi^{k-1}(z_1)$
over~$F$. So we need only to compute hyperexponential solutions
of~$L(z_1)=0$. If $k<n$, then we compute hyperexponential solutions
of~(\ref{EQ:sys1}), in which~$z_1 \neq 0$ and~$z_1=0$, separately.
In the former case, let~$h$ be a hyperexponential solution
of~$L(z_1)=0$, then, all hyperexponential solutions
of~(\ref{EQ:sys1}) of the form~$h(v_1, \ldots, v_n)^T$ can be found
by substituting~$hZ$ into~(\ref{EQ:sys1}) and computing the rational
solutions of the resulting equation. There are methods for computing
rational solutions of linear functional matrix equations
in~\cite{AbramBronst01, Barkatou99differential}. In the latter
case~$z_1=0$, we compute~$P, Q$ and a partition of~$(z_2, \ldots,
z_n)^T$ into two sub-vectors $Y_1$ and~$Y_2$ such that~$\phi(Y_1)=P
\, Y_1$ and~$Y_2=Q \, Y_1$, by an ordinary version of the algorithm
LinearReduction described in \cite[Section 2.5.3]{Wu_thesis}. Then
we apply the same method to~$\phi(Y_1)=P \, Y_1$, recursively.

In Section~\ref{SUBSECT:partial}, one will encounter a matrix
equation of form~$V\phi(Y){=}UY$ where~$Y$ is a vector of
unknowns,~$U$ and~$V$ are matrices over~$F$, and~$V$ has full column
rank. A similar reduction transforms the equation
into~$\{\,\phi(Y_1)=U^\prime Y_1,$ \, $ Y_2=V^\prime Y_1\}$,
where~$Y_1$ and~$Y_2$ form a partition of~$Y$ into two sub-vectors
of unknowns, and $U^\prime$ and $V^\prime$ are some matrices
over~$F$. So we can find hyperexponential solutions
of~$V\phi(Y)=UY$.

\begin{ex} \label{EX:ordinary}

Let~$F=\bC(x,m,n)$  and~$\sigma_n$ be the shift operator with
respect to~$n$. We now compute hyperexponential solutions of the
matrix difference equation~$\cA: \sigma_n(Z){=} A Z$ where~$Z=(z_1,
z_2, z_3)^T$ and {\small $$ A{=}\mskip-8mu\left(\mskip-14mu
\begin{array}{ccc} \frac{n(2nx+x-2x^2-1)}{2(nx-1)}& \frac{x(-n-3+2x+2nx)}{2(nx-1)}&
0 \smallskip\\
\frac{n(n-1-x+nx)}{2(nx-1)}& \frac{-2n-2+x+2nx+n^2x}{2(nx-1))}
&0\smallskip\\
\frac{n^2x+3nx+2nm^2-n^2-n+2m^2}{2(nx-1)}&
\frac{x+2m^2-n^2x+2xm^2+2x^2n}{2(1-nx)}& x\\
\end{array}\mskip-12mu\right)\mskip-8mu.
$$}

By linear algebra, we find a linear difference equation:
$L(z_1)=\sigma_{n}^2(z_1)
{+}\frac{4x^2n{+}11{-}18nx{-}6n^2x{-}12x{+}14n{+}3n^2{+}4x^2}{2({-}n{-}3{+}2x{+}2nx)}\sigma_n(z_1)$
$-\frac{n(8x^2+4-12x-8nx+4x^2n-2n^2x+5n+n^2)}{2(-n-3+2x+2nx)}z_1=0\,.
$ All hyperexponential solutions of~$L$ are of the form~$c\Gamma(n)$
for $c\in \bC(x,m)$. Substituting~$Z=hY$ with~$h=\Gamma(n)$
and~$Y=(y_1, y_2, y_3)^T$ into~$\cA$, we get
a~$\bC(x,m)$-basis~$V=\left(\frac{n+1}{x}, \frac{(1+x)n}{x^2},
\frac{nx+m^2}{x^2}\right)^T$ of rational solutions of the resulting
system. So~$\{(\Gamma(n), V) \}$ is among the representation of
hyperexponential solutions of~$\cA$. In addition,~$z_1$ is not a
cyclic vector as the order of~$L$ is less than the size of~$\cA$. By
substituting~$z_1=0$ into~$\cA$ we get~$z_1=0$, $z_2 = 0$
and~$\sigma_n(z_3) = xz_3.$ Thus~$z_3 = c\,x^n$ for any~$c\in
\bC(x,m)$, hence~$c\,x^n(0, 0, 1)^T$ is a hyperexponential solution
of~$\cA$. So  a representation of hyperexponential solutions
of~$\cA$ is~$\left\{\, \left(\Gamma(n), \, V\right), \,\, \left(x^n,
\, (0, 0, 1)^T\right)\, \right\}.$
\end{ex}

\subsection{Partial case} \label{SUBSECT:partial}
Let~$|\Delta^\prime|{=}m-1$ and $\Delta \setminus \Delta^\prime
=\{\phi\}$. Assume that~$ \{\,(h_1^\prime, \, V_1^\prime)$, $
\ldots,\,\, (h_{t}^\prime, V_{t}^\prime)\,\} $
 is a representation for
hyperexponential solutions of the system associated to $(M,
\Delta^\prime)$. We decide whether~$h_1^\prime, \ldots,
h_{t}^\prime$ are extensible for~$\phi$. If none of them is
extensible, then the system~$\cA$ associated to~$(M, \Delta)$ has no
hyperexponential solution by the proof of
Proposition~\ref{PROP:ext}. Otherwise, we may further assume
that~$h_1^\prime, \ldots, h_s^\prime$ are extensible to~$h_1,
\ldots, h_s$, respectively, while~$h_{s+1}^\prime,$ \ldots,
$h_t^\prime$ are not extensible. By Propositions~\ref{PROP:ext}, for
every hyperexponential solution of~$\cA$, there exists a unique~$k$
in~$\{1, \ldots, s\}$ such that the solution is of the
form~$h_kV_k^\prime D_k$, where~$D_k$ is a hyperexponential vector
with constant entries with respect~to~$\Delta^\prime$.

Let~$\phi(Z)=BZ$ be the equation corresponding to~$\phi$ in~$\cA$.
For~$1 \le k\le s$, substituting~$h_k V_k^\prime D_k$
into~$\phi(Z)=BZ$ yields an equation~$Q_k\phi(D_k) = B_kD_k$ for
some matrices~$Q_k,\,B_k$ over~$F$. In addition,~$Q_k$ has full
column rank. As in the proof of Lemma~\ref{LM:ext}, we choose a
$C^\prime(F)$-basis~$\{\alpha_i\}$ of~$F$, and write~$Q_k = \sum_i
Q_{ki} \alpha_i$ and~$B_k=\sum_{i} B_{ki} \alpha_i$, where~$Q_{ki}$
and~$B_{ki}$ are matrices over~$C^\prime(F)$. Let~$U_k$ and~$W_k$ be
matrices formed by the stacking of the non-zero matrices~$Q_{ki}$
and~~$B_{ki}$, respectively. By Lemma~\ref{LM:wr-cas}
\begin{equation} \label{EQ:sys2}
{U_k \phi(D_k)= W_k D_k,}
\end{equation}
where~$U_k$ has full column rank since $Q_k$ has. We compute
hyperexponential solutions of~(\ref{EQ:sys2}) over~$C^\prime(F)$.

Assume that, for~$1 \le k \le l$,~$ \{(g_{k1}, G_{k1}),\ldots,
(g_{ki_k}, G_{ki_k})\}$ is a representation of hyperexponential
solutions of~(\ref{EQ:sys2}), while~(\ref{EQ:sys2}) has no
hyperexponential solutions for any~$k$ with $l<k \le s$. Then a
representation of hyperexponential solutions of~$\cA$ consists
of~$(f_{kj}, \, V_{kj})$, where~$f_{kj}=g_{kj}h_k$, the set of the
column vectors of~$V_{kj}$ is a maximal set of linearly independent
column vectors of the matrix $V^\prime_k G_{kj}$, $j=1, \ldots,
i_k$, and~$k=1, \ldots, l$. To prove this assertion, we need only to
show that~a hyperexponential solution of~$\cA$ cannot be represented
by both~$(f_{kj}, V_{kj})$ and~$(f_{kj^\prime}, V_{kj^\prime})$
with~$j \neq j^\prime$. Suppose the contrary, then there exists~$r
\in F$ such that~$rf_{kj}=f_{kj^\prime}$. It follows
that~$rg_{kj}=g_{kj^\prime}$, so $(g_{kj}, G_{kj})$ and
$(g_{kj^\prime}, G_{kj^\prime})$ would also represent the same set
of hyperexponential vectors, a contradiction.

We illustrate the algorithm by two examples. The first one cannot
be handled directly by the method in~\cite{LabahnLi04}.

\begin{ex} \label{EX:par1}
Consider the field $F {=} C(x,y)$ with $\Delta {=} \{\delta,
\sigma\}$ where~$C=\bQ(e)$, $\delta = \frac{\partial}{\partial x} +
\frac{\partial}{\partial y}$ and $\sigma$ is defined by $\sigma(x) =
x+1$ and~$\sigma(y) = y$.  The constants of~$F$ are $C$, as the
constants with respect to~$\sigma$ are~$C(y)$ and the constants
of~$C(y)$ with respect to~$\delta$  are $C$.
 Let us compute
hyperexponential solutions of the system~$\{\,\sigma(Z)= A_s Z, \,
\delta (Z)= A_d Z\} $ where $Z=(z_1, z_2, z_3, z_4)^T$,
$$
{A_s = \left(\begin{array}{cccc} 0 & \frac{1}{y} & -x\, e & e+1 \\
-y\,e & e+1& 0& y\,e \\ 0 & 0& 0 & \frac{1}{x+1} \\
0 & 0& -x\,e & e+1
\end{array}\right)}
$$
and
$$
A_d= \left(\begin{array}{cccc} -\frac{1}{y} & -\frac{4}{2y-1} &
\frac{x}{y} & \frac{2y-1+4y^2}{y(2y-1)} \\ 0 & 0& 0& 1\\
-\frac{4y}{x(2y-1)} & 0& \frac{4yx-2y+1}{x(2y-1)}&
\frac{4y}{x(2y-1)}\\ 0 & -\frac{4}{2y-1} & 0 & \frac{4y}{2y-1}
\end{array}\right).
$$

We obtain a representation~$\left\{\left(h_1=1, V_1\right),\,
\left(h_2=e^x,V_2\right)\right\} $ of hyperexponential solutions of
the first matrix equation $\sigma(Z)=A_sZ$ with
$$
V_1  =\left(\begin{array}{cc}
\frac{1}{y} & 1\\
1 & 0 \\
0 & \frac{1}{x} \\
0 & 1
\end{array}\right), \quad V_2 = \left(
\begin{array}{cc}
\frac{1}{y\,e} & 1 \\
1 & 0 \\
0 & \frac{1}{x\,e}\\
0 & 1 \end{array} \right).
$$
Both~$h_1$ and~$h_2$ are extensible for~$\delta$. Suppose
that~$h_1 V_1D$ is a solution of~$\cA$ for some hyperexponential
vector~$D$ over~$C(y)$. To decide~$D$, substitute~$h_1V_1D$
for~$Z$ into the second matrix equation~$\delta(Z)= A_dZ$ to yield
$$
\left(\begin{array}{cc}
\frac{1}{y} & 1\\
1 & 0 \\
0 & \frac{1}{x} \\
0 & 1
\end{array}\right) \delta (D) = \left(\begin{array}{cc} -\frac{4}{2y-1} &
\frac{2y-1+4y^2}{y(2y-1)} \\ 0 & 1 \\ -\frac{4}{x(2y-1)}&
\frac{4y}{x(2y-1)}  \\ -\frac{4}{2y-1} & \frac{4y}{2y-1}
\end{array}\right) D,
$$
which translates to a matrix equation of size two {\small $$
\delta (D) = \left(\begin{array}{cc} 0 & 1 \\
-\frac{4}{2y-1} & \frac{4y}{2y-1} \end{array}\right) D.
$$}
A representation of hyperexponential solutions of the above system
is $ \left\{\, (1,\, U_1 := (y,1)^T), \,(\, e^{2y}, U_2 := (1,
2)^T)\,\right\}. $ Hence the original system has hyperexponential
solutions given by $\{(1, V_1U_1),$ $ (e^{2y}, V_1U_2)\}$.

Similarly, substituting~$h_2V_2D$ for~$Z$ into~$\delta(Z)= A_dZ$
finally yields
a matrix equation of size two {\small $$
\delta (D) = \left(\begin{array}{cc} -1 & 1 \\
-\frac{4}{2y-1} & \frac{2y+1}{2y-1} \end{array}\right) D.
$$}
A representation of hyperexponential solutions of the above system
is~$ \left\{\left(e^y, W_1:=(e, 2e)^T\right),  \left(e^{-y},
W_2:=(y\, e, e)^T\right)\right\}.$ So the original system has
hyperexponential solutions given by $\{(e^{x+y}, V_2W_1),\,(e^{x-y},
V_2W_2)\} $. Accordingly,
\begin{eqnarray*}
(1,\, (2, y, \frac{1}{x}, 1)^T), & (e^{2y}, (\frac{1}{y}+2,
1,\frac{2}{x}, 2)^T), \\
(e^{x+y}, (\frac{1}{y}+2e, e, \frac{2}{x}, 2e)^T),& (e^{x-y}, (1+e,
ye, \frac{1}{x}, e)^T)
\end{eqnarray*}
form a representation of hyperexponential solutions of the original
system.
\end{ex}
\begin{ex} \label{EX:par2}
Let~$F=\bC(x, y, k)$, and~$\delta_x,\delta_y$ and~$\sigma_k$ denote
partial differentiations with respect to~$x, y$ and the shift
operator with respect to~$k$, respectively.
Let~$\LO{=}F[\pa_x,\pa_y, \pa_k, \pa_k^{-1}]$ be the Laurent-Ore
algebra over~$F$ and~$M$ be an~$\LO$-module with
an~$F$-basis~$\{\ve_1,\ve_2, \ve_3\}$ whose structure matrices are
$-A_x^T$, $(A_k^{-1})^T $ and~$-A_y^T$ where {
$$ A_x {=} \mskip-5mu\left(\mskip-10mu\begin{array}{ccc}
\frac{x+y}{xy} &
-\frac{k(2x+k)}{x(x+k)} & 0 \smallskip \\
0 & \frac{-y+x+k}{y(x+k)} & 0\smallskip \\
\frac{3x+2y}{x+y} & -\frac{k(3x+2y)}{x+y} & \frac{x}{y(x+y)}
\end{array}\mskip-10mu\right)\mskip-5mu,
$$
$$
 A_k {=} \mskip-5mu\left(\mskip-10mu\begin{array}{ccc} \frac{k(y+k)}{y+k+1} &
\frac{k(k^2+2xk+xy+x+k)}{(y+k+1)(x+k+1)}& 0 \smallskip \\
 0 &
\frac{k(x+k)}{x+k+1} & 0\smallskip \\
-\frac{x(2k+y+1)}{y+k+1} & \frac{xk(2k+y+1)}{y+k+1} & k+1
\end{array}\mskip-10mu\right)\mskip-5mu,
$$}
and
$$
A_y = \left(\begin{array}{ccc} -\frac{y^2+xy+xk}{(y+k)y^2} &
\frac{k(2y+k)}{y(y+k)}& 0 \smallskip \\
0 & -\frac{x-y}{y^2} & 0 \smallskip \\
-\frac{x(2xy+y^2+xk)}{y(y+k)(x+y)} &
\frac{xk(2xy+y^2+xk)}{y(y+k)(x+y)} & -\frac{x^2}{y^2(x+y)}
\end{array}\right).
$$

We compute all hyperexponential solutions of the associated
system~$\cA: \{\delta_x(Z){=} A_x Z, \sigma_k(Z)= A_k Z,  \delta_y
(Z)= A_y Z\}$ of~$M$ where~$Z=(z_1, z_2, z_3)^T$. {A representation
of hyperexponential solutions of~$\delta_x(Z)=A_xZ$
is~$\{(e^{\frac{x}{y}}, V)\}$ where} {\small $$
V=\left(\begin{array}{ccc}\frac{k}{x+k}& 0 & x \smallskip \\
\frac{1}{x+k}& 0 & 0 \smallskip \\
0& \frac{1}{x+y} &   x^2 \end{array}\right).
$$}

Clearly,~$h=e^{\frac{x}{y}}$ is extensible for~$\sigma_k$. {Suppose
that~$h V D$ is a solution of $ \{\,\delta_x(Z){=}A_x Z$,
$\sigma_k(Z)= A_kZ\}$} for some hyperexponential vector~$D$ over
$\bC(k,y)$ with respect to~$\{\delta_x, \sigma_k\}$. To
identify~$D$, substitute $hVD$ into the second matrix
equation~$\sigma_k(Z)=A_k Z$ to yield {\small
\begin{equation*}
\left(\mskip-7mu\begin{array}{ccc} \frac{k+1}{x+k+1}& 0& x\smallskip \\
\frac{1}{x+k+1} & 0 & 0\smallskip \\
0 & \frac{1}{x+y} & x^2\end{array}\mskip-7mu\right) \sigma_k(D) {=}
\mskip-2mu\left(\mskip-7mu\begin{array}{ccc} \frac{k(k+1)}{x+k+1} &
0 &
\frac{k(y+k)x}{y+k+1} \smallskip \\
\frac{k}{x+k+1} & 0 & 0\smallskip \\ 0 & \frac{k+1}{x+y} &
\frac{x^2k(y+k)}{y+k+1}\end{array}\mskip-7mu\right)\mskip-5muD,
\end{equation*}}
which translates to the system
\begin{equation*}
{\sigma_k(D)= \left(\begin{array}{ccc} k & 0& 0\\ 0 & k+1& 0\\
0 & 0& \frac{k(y+k)}{y+k+1}\end{array}\right) D.}
\end{equation*}
Its hyperexponential solutions are {given} 
by~$\{(\Gamma(k),U )\}$ where {
 $$ U= \left(\begin{array}{ccc} 1 &
0&0
\smallskip \\ 0 & k &0\smallskip \\0 & 0 & \frac{1}{y+k}\smallskip
\end{array}\right).
$$} Hence hyperexponential solutions of
the first two matrix equations have a
representation~$\{(e^{\frac{x}{y}}\Gamma(k),  VU)\}$.

Carrying on the above process, we  find a representation
$\{(e^{\frac{x}{y}}\Gamma(k), \, W)\}$ of hyperexponential solutions
of the original system where
$$
W=\left(\begin{array}{ccc}\frac{ky}{x+k} & 0 & \frac{x}{y+k} \\
\frac{y}{x+k} &0 & 0\\
0 &   \frac{ky}{x+y} & \frac{x^2}{y+k}\end{array}\right).
$$
So~$\{W\}$ is  a representation of one-dimensional submodules of~$M$
relative to the given basis. For this example,~$M$ is a direct sum~$
F\vw_1 \oplus F\vw_2 \oplus F\vw_3 $ where~$\vw_1=
\frac{ky}{x{+}k}\ve_1{+} \frac{y}{x{+}k}\ve_2$, $\vw_2 =
\frac{ky}{x{+}y}\ve_3$ and~$\vw_3 = \frac{x}{y{+}k}\ve_1 {+}
\frac{x^2}{y{+}k}\ve_3$.
\end{ex}

To a finite-dimensional linear functional system, one can
associate a fully integrable system. Proposition 2
in~\cite{BronsteinLiWu05}  describes a one-to-one correspondence
between the solutions of the given system and those of the
associated one (see also  Proposition 2.4.12 in \cite{Wu_thesis}).
Consequently, the algorithm in this section can be used for
computing hyperexponential solutions of finite-dimensional linear
functional systems.



\begin{thebibliography}{10}

\bibitem{AbramBronst01}
S.~Abramov and M.~Bronstein.
\newblock On solutions of linear functional systems.
\newblock In B.~Mourrain, ed.\, {\em Proc. ISSAC'2001}, pp.\ 1--6. ACM
  Press, 2001.

\bibitem{AbrPaulePetk98}
S.~Abramov, P.~Paule, and M.~Petkov\v{s}ek.
\newblock q-hypergeometric solutions of q-difference equations.
\newblock {\em Discrete Math.}, 180:3--22, 1998.

\bibitem{Barkatou99differential}
M.~Barkatou.
\newblock On rational solutions of systems of linear differential equations.
\newblock {\em J. Symbolic Comput.}, 28(4/5):547--568, 1999.

\bibitem{BarkPflueg99}
M.~Barkatou and E.~Pfl{\"u}gel.
\newblock An algorithm computing the regular formal solutions of a system of
  linear differential equations.
\newblock {\em J. Symbolic Comput.}, 28:569--587, 1999.

\bibitem{Beke1894}
E.~Beke.
\newblock Die irreducibilit{\"a}t der homogenen differentialgleichungen.
\newblock {\em Math.~Annal.}, 45:278--294, 1894.

\bibitem{Bronstein92linear}
M.~Bronstein.
\newblock Linear ordinary differential equations: breaking through the order 2
  barrier.
\newblock In P.~S. Wang, ed.\, {\em Proc. ISSAC'1992}, pp.\ 42--48. ACM
  Press, 1992.

\bibitem{BronsteinLiWu05}
M.~Bronstein, Z.~Li, and M.~Wu.
\newblock Picard--{V}essiot extensions for linear functional systems.
\newblock In M.~Kauers, ed.\, {\em Proc. ISSAC'2005}, pp.\ 68--75. ACM
  Press, 2005.

\bibitem{GrSch05}
D.~Grigoriev and F.~Schwarz.
\newblock Generalized {L}oewy-decomposition of ${D}$-modules.
\newblock In M.~Kauers, ed.\, {\em Proc. ISSAC'2005}, pp.\ 163--170. ACM
  Press, 2005.

\bibitem{vanHoeij97}
M.~van Hoeij.
\newblock Factorization of differential operators with rational functions
  coefficients.
\newblock {\em J. Symbolic Comput.}, 24(5):537--561, 1997.

\bibitem{vanHoeij99}
M.~van Hoeij.
\newblock Finite singularities and hypergeometric solutions of linear
  recurrence equations.
\newblock {\em Journal of Pure and Applied Algebra}, (139):109--131, 1999.

\bibitem{LabahnLi04}
G.~Labahn and Z.~Li.
\newblock Hyperexponential solutions of finite-rank ideals in orthogonal {O}re
  algebras.
\newblock In J.~Gutierrez, ed.\, {\em Proc. ISSAC'2004}, pp.\ 213--220. ACM
  Press, 2004.

\bibitem{Lang}
S.~Lang.
\newblock {\em Algebra, Graduate Texts in Mathematics}, volume 211.
\newblock Springer, 2002.

\bibitem{LiSchwarz01}
Z.~Li and F.~Schwarz.
\newblock Rational solutions of {R}iccati-like partial differential equations.
\newblock {\em J. Symbolic Comput.}, 31:691--716, 2001.

\bibitem{LiSchwarzTsarev03}
Z.~Li, F.~Schwarz, and S.~Tsarev.
\newblock Factoring systems of linear {PDE}'s with finite-dimensional solution
  spaces.
\newblock {\em J. Symbolic Comput.}, 36:443--471, 2003.

\bibitem{Petk92}
M.~Petkov{\v{s}}ek.
\newblock Hypergeometric solutions of linear recurrences with polynomial
  coefficients.
\newblock {\em J. Symbolic Comput.}, {\bf 14}:243--264, 1992.

\bibitem{VdpSinger03}
M.~van der Put and M.F.~Singer.
\newblock {\em Galois Theory of Linear Differential Equations, Grundlehren der
  Mathematischen Wissenschaften}, volume 328.
\newblock Springer, Heidelberg, 2003.

\bibitem{Singer91}
M.F.~Singer.
\newblock Liouvillian solutions of linear differential equations with
  liouvillian coefficients.
\newblock {\em J. Symbolic Comput.}, {\bf 11}:251--273, 1991.

\bibitem{Wu_thesis}
M.~Wu.
\newblock {\em On Solutions of Linear Functional Systems and Factorization of
  Modules over {L}aurent-{O}re Algebras}.
\newblock PhD thesis, Academia Sinica and Universit\'e de Nice,
  \url{http://www.mmrc.iss.ac.cn/~mwu/Thesis/Wu-thesis.pdf}, 2005.

\end{thebibliography}
\balancecolumns
\end{document}